\input phyzzx
\hsize=417pt 
\sequentialequations
\Pubnum={ EDO-EP-12}
\date={ \hfill April 1997}
\titlepage
\vskip 32pt
\title{ Quantum Gravity near Apparent Horizon and Two Dimensional Dilaton 
Gravity }
\vskip 12pt
\author{Ichiro Oda \footnote\dag {E-mail address: 
ioda@edogawa-u.ac.jp}}
\vskip 12pt
\address{ Edogawa University,                                
          474 Komaki, Nagareyama City,                        
          Chiba 270-01, JAPAN     }                          
%
%
%
%
%
\abstract{ We study the Hawking radiation in two dimensional dilaton 
black hole by means of quantum gravity holding near the apparent 
horizon. First of all, we construct the canonical formalism of the 
dilaton gravity in two dimensions. Then the Vaidya metric corresponding 
to the dilaton black hole is established where it is shown that the 
dilaton field takes a form of the linear dilaton. Based on the canonical 
formalism and the Vaidya metric, we proceed to analyze quantum properties 
of a dynamical black hole. It is found that the mass loss rate of the 
Hawking radiation is independent of the black hole mass and at the same 
time the apparent horizon recedes to the singularity as shown in other 
studies of two dimensional gravity. It is interesting that one can 
construct quantum gravity even near the curvature singularity and draw 
the same conclusion with respect to the Hawking radiation as the 
above-mentioned picture. Unfortunately, the present formalism seems to be 
ignorant of the contributions from the functional measures over the 
gravitational field, the dilaton and the ghosts.
\endpage
%
%
%

\def\sp(#1){\noalign{\vskip #1pt}}

%
%
%
%
%
\topskip 30pt
\par
\leftline{\bf 1. Introduction}	
\par
More than twenty years ago, Hawking [1] has shown that black holes are 
not completely black and emit thermal radiation with a definite 
temperature through quantum mechanical pair creation of particles near 
the horizon in a gravitational field where one member of the pair drops 
in a black hole while the other escapes to infinity. This result was 
derived in the context of ``semiclassical'' approach, where the effects 
of gravitation are still represented by a classical spacetime $(M, 
g_{ab})$, while matter fields are treated as quantum fields propagating 
in this classical spacetime. Subsequent investigations have been 
focused on understanding of serious problems raised by the Hawking 
radiation, concerning the fate of quantum information [2] and the 
statistical mechanical picture of black hole thermodynamics [3] e.t.c.

In recent papers, a quantum formalism has been proposed for the study of 
black hole quantum mechanics [4, 5]. The critical idea behind this 
formalism is that some essential features of quantum black holes might be 
intimately related to the quantum mechanical behavior of the black hole 
horizon, thus it might be sufficient to establish the quantum gravity 
holding particularly near the horizon to understand an overall picture 
of quantum black holes. Indeed, afterward, this formalism has been 
fruitfully applied to several problems associated with quantum black 
holes in three [6] and four [7] spacetime dimensions. 

In the course of applications, we have wondered to what extent the 
quantum gravity near the horizon would describe quantum aspects of a black 
hole. To address this question, it is tempting to try to apply the 
formalism to well-understood model of quantum black holes, that is, the 
dilaton gravity in 1+1 dimensions (CGHS model) [8]. The dilaton gravity 
in two dimensions enjoys nice features of black-hole formation and/or 
evaporation shared with the spherically symmetric black hole in 3+1 
dimensions. Therefore, this toy model has raised hopes that a 
satisfactory description of black hole evolution might be accounted for 
in a very simplified setting.

The article is organized as follows. In section 2, we construct the 
canonical formalism of the two dimensional dilaton gravity. 
In section 3, we derive the Vaidya metric corresponding to the dilaton black 
hole. The canonical formalism and the Vaidya metric are used to construct 
a quantum theory holding in the vicinity of the apparent horizon of the 
dilaton black hole in section 4.  In section 5, we analyse the Hawking 
radiation from purely quantum mechanical viewpoint. Here it is shown that 
the mass loss rate is independent of black hole mass. The last section is 
devoted to conclusion.

\vskip 32pt
\leftline{\bf 2. ADM Canonical Formalism of Two Dimensional Dilaton 
Gravity }	
\par
We begin our investigations by constructing the ADM first-order canonical 
formalism of two dimensional dilaton gravity.

The action that we start with has the well-known form [8]
$$ \eqalign{ \sp(2.0)
S = {1 \over 2G} \int  d^2 x \sqrt{-g} \ e^{-2 \phi}  \ \bigl[ R + 4 
(\nabla \phi)^2 + 4 \lambda^2 \bigr] - {1 \over 2} \int  d^2 x 
\sqrt{-g} \ (\nabla f)^2,
\cr
\sp(3.0)} \eqno(1)$$
with the dilaton field $\phi$, the cosmological constant $\lambda$, and 
the single massless conformal matter field $f$. Differing from the 
original CGHS convention [8] where $G = \pi$, we will take $G = {1 \over 
2}$ in this paper. Related to this choice, we have modified 
the coefficient in front of the matter action from the CGHS value $- 
{1 \over {4 \pi}}$ to $- {1 \over 2}$, which is natural from the viewpoint 
of the spherically symmetric reduction of the four dimensional gravity [9]. 

Let us adopt the ADM splitting of 1+1 dimensional spacetime given by
$$ \eqalign{ \sp(2.0)
g_{ab} = \left(\matrix{ { - \alpha^2 + {\beta^2 \over \gamma}} & \beta \cr
              \beta & \gamma \cr} \right).
\cr
\sp(3.0)} \eqno(2)$$
Then the normal unit vector $n^a$ orthogonal to the hypersurfaces 
$x^0 = const$ reads 
$$ \eqalign{ \sp(2.0)
n^a = ( {1 \over \alpha}, \ - {\beta \over {\alpha \gamma}}),
\cr
\sp(3.0)} \eqno(3)$$
and the projection operator $h^{ab}$ over the $x^0 = const$ hypersurfaces 
becomes  
$$ \eqalign{ \sp(2.0)
h^{ab} &= g^{ab} + n^a n^b,  
\cr
       &= \left(\matrix{ 0 & 0 \cr
               0 & 1 \over \gamma \cr} \right).
\cr
\sp(3.0)} \eqno(4)$$
In terms of the ADM parametrization (2), after some calculations the 
action (1) can be written as
$$ \eqalign{ \sp(2.0)
S &= \int d^2x \ L,  
\cr
&= \int d^2x \ \bigl[ \ 4 \alpha \sqrt \gamma \ e^{-2 \phi} \ \bigl\{ 
\lambda^2  - (n^a \partial_a \phi)^2 + {1 \over \gamma} (\phi^\prime)^2 + 
K n^a \partial_a \phi - {\alpha^\prime \over {\alpha \gamma}} \phi^\prime 
\bigr\}  
\cr
&\qquad+   {1 \over 2} \alpha \sqrt \gamma \ \bigl\{{ ( n^a 
\partial_a f )^2 - {1 \over \gamma} ( f^\prime )^2  }\bigr\} \bigr],
\cr
\sp(3.0)} \eqno(5)$$
where the trace of the extrinsic curvature $K = g^{ab} K_{ab}$ is
$$ \eqalign{ \sp(2.0)
K = {1 \over \sqrt{-g}} \partial_a ( \sqrt{-g} \ n^a ) 
= {\dot \gamma \over {2\alpha\gamma}} - {\beta^\prime \over 
{\alpha\gamma}} + {\beta \over {2\alpha\gamma^2}} \gamma^\prime,
\cr
\sp(3.0)} \eqno(6)$$
and ${\partial \over {\partial x^0}} = \partial_0$ and ${\partial \over 
{\partial x^1}} = \partial_1$ are also denoted by an overdot and a prime, 
respectively. In deriving (5), we have used the formula [9, 10]
$$ \eqalign{ \sp(2.0)
R = 2 n^a \partial_a K + 2 K^2 - {2 \over {\alpha \sqrt \gamma}} 
( { \alpha^\prime \over \sqrt \gamma} )^\prime.
\cr
\sp(3.0)} \eqno(7)$$

The action (5) indicates that $\alpha$ and $\beta$ are non-dynamical 
Lagrange multiplier fields due to the absence of the term including 
$x^0$-differentiation so we regard the massless matter field $f$, the 
dilaton field $\phi$ and the ``graviton'' $\gamma$ as the dynamical 
fields. Then canonical conjugate momenta can be read off from the action 
(5):
$$ \eqalign{ \sp(2.0)
p_f &= \sqrt \gamma \ n^a \partial_a f,
\cr
p_{\phi} &= 4 \sqrt \gamma \ e^{-2 \phi} \ ( -2 n^a \partial_a \phi + K 
), 
\cr
p_{\gamma} &= {2 \over \sqrt \gamma} \ e^{-2 \phi} \ n^a \partial_a \phi.
\cr
\sp(3.0)} \eqno(8)$$
Now it is straightforward to derive the Hamiltonian whose result is given 
by 
$$ \eqalign{ \sp(2.0)
H &= \int dx^1 \ ( p_f \dot f + p_{\phi} \dot \phi + p_{\gamma} \dot 
\gamma - L ),  
\cr
  &= \int dx^1 \ ( \alpha H_0 + \beta H_1  ), 
\cr
\sp(3.0)} \eqno(9)$$
where the constraints are explicitly of form
$$ \eqalign{ \sp(2.0)
H_0 &= {1 \over {2 \sqrt{\gamma} }} \ p_f ^2 - 4 \sqrt \gamma \ e^{-2 \phi} 
\ \lambda^2 - {4 \over \sqrt \gamma} \ e^{-2 \phi} \ (\phi^\prime)^2  
- ( {4 \over \sqrt \gamma} \ e^{-2 \phi} \ \phi^\prime )^\prime  
\cr
&\qquad+ {1 \over { 2 \sqrt \gamma}} \ ( f^\prime )^2 + {\sqrt \gamma 
\over 2} \ e^{+2 \phi} \ p_\phi p_\gamma + \gamma \sqrt \gamma \ e^{+2 
\phi} \  p_\gamma ^2 , 
\cr
\sp(3.0)} \eqno(10)$$
$$ \eqalign{ \sp(2.0)
H_1 = {1 \over \gamma} \ p_f f^\prime + {1 \over \gamma} p_\phi 
\phi^\prime - 2  p_\gamma ^\prime - {1 \over 
\gamma} p_\gamma \gamma^\prime. 
\cr
\sp(3.0)} \eqno(11)$$
Note that $H_0$ and $H_1$ are generators corresponding to the time 
translation and the spatial displacement, respectively. Also let us notice 
that $\alpha$ and $\beta$ are certainly the Lagrange multiplier fields as 
mentioned before. At this stage, it is easy to derive the ADM surface 
term via the dual Legendre transformation by following Regge and 
Teitelboim [11], though we now omit the detail since it is not so much 
important for later discussions.

\vskip 32pt
\leftline{\bf 3. Vaidya Metric}	
\par
In this section, we will construct the Vaidya metric to the two 
dimensional dilaton black hole. This Vaidya metric will be used in later 
sections when we wish to discuss the Hawking radiation arising from a 
dynamical black hole.

As a simple illustration, let us recall how to build the Vaidya metric 
to the Schwarzschild black hole in four dimensions. Neglecting the 
irrelevant  
angular parts $(\theta, \varphi)$, the Schwarzschild geometry has the 
famous form
$$ \eqalign{ \sp(2.0)
ds^2 = - (1 - {2M \over r}) dt^2 + {1 \over {1 - {2M \over r}}} dr^2.
\cr
\sp(3.0)} \eqno(12)$$
Introducing the advanced time coordinate $v = t + r^*$ with the tortoise 
coordinate $dr^* = {dr \over {-g_{00}}}$, in the $(v, r)$ 
coordinates the Schwarzschild metric can be transformed to
$$ \eqalign{ \sp(2.0)
ds^2 = - (1 - {2M \over r}) dv^2 + 2 dv dr.
\cr
\sp(3.0)} \eqno(13)$$
A generalization of a constant mass $M$ to the mass function $M(v)$ 
gives rise to the Vaidya metric corresponding to the Schwarzschild black 
hole. The reason why we prefer the Vaidya metric to the Schwarzschild one 
is that the former satisfies the classical field equations as it is when 
there is a flow of matters with a form of the energy-momentum tensor 
$T(v)$, while in the latter the mass function must have a complicated 
dependence on the coordinates to satisfy them, which makes the following 
analysis ugly. In other words, the Vaidya form of a black hole provides 
us a convenient playground to discuss the properties of a dynamical black 
hole.  

In order to have a close relationship with CGHS work [8], let us start by 
their black hole solution in the lightcone coordinates
$$ \eqalign{ \sp(2.0)
ds^2 &= - e^{2 \rho} dx^{+} dx^{-},
\cr
     &= - {1 \over {{M \over \lambda} - \lambda^2 x^{+} x^{-}}} dx^{+} 
     dx^{-}. 
\cr
\sp(3.0)} \eqno(14)$$
Transforming the coordinates from the $(x^{+}, x^{-})$ to the $(v, r)$ 
which are related to each other
$$ \eqalign{ \sp(2.0)
x^{+} &= {\sqrt M \over {\lambda \sqrt \lambda}} e^{\lambda v},
\cr
x^{-} &= - {1 \over {\sqrt {\lambda M}}} (e^{2 \lambda r} - {M \over 
\lambda}) e^{- \lambda v}, 
\cr
\sp(3.0)} \eqno(15)$$
the two dimensional line element (14) reduces to
$$ \eqalign{ \sp(2.0)
ds^2 = - (1 - {M \over \lambda} e^{- 2 \lambda r}) dv^2 + 2 dv dr.
\cr
\sp(3.0)} \eqno(16)$$
Then the Vaidya metric corresponding to the dilaton black hole can be 
obtained by promoting a constant mass to the mass function $M(v)$ 
depending on only the $v$ coordinate. Here it is worthwhile to notice that 
in the newly introduced coordinates $(v, r)$, the dilaton field which was 
given by $\phi = \rho$ in the lightcone coordinates $(x^{+}, x^{-})$, 
takes a remarkably simple form, that is, a linear dilaton form 
$$ \eqalign{ \sp(2.0)
\phi = - \lambda r.
\cr
\sp(3.0)} \eqno(17)$$
This would lead to a great advantage in analyzing the constraints as 
well as the field equations in the below.

Indeed, it is verified that the solutions (16) and (17) are an extremum 
of the action (1) near the apparent horizon
$$ \eqalign{ \sp(2.0)
r_{AH} = - {1 \over {2 \lambda}} \log {\lambda \over M},
\cr
\sp(3.0)} \eqno(18)$$
whose definition arises from the condition $g_{vv} = 0$ which is also 
consistent with the usual definition $(\nabla \phi)^2 = 0$ in the two 
dimensional dilaton gravity. The classical field equations are easily 
obtained from the action (1):
$$ \eqalign{ \sp(2.0)
&{} 2 e^{-2 \phi} \ \bigl[\nabla_a \nabla_b \phi + g_{ab} \ 
\bigl\{(\nabla \phi)^2 - \nabla^2 \phi - \lambda^2 \bigr\} \bigr]
\cr
&\qquad= {1 \over 2} \bigl[ \nabla_a f  \nabla_b f - {1 \over 2} g_{ab} ( 
\nabla f)^2 \bigr],
\cr
\sp(3.0)} \eqno(19)$$
$$ \eqalign{ \sp(2.0)
R + 4 \lambda^2 + 4 \nabla^2 \phi - 4 (\nabla \phi)^2  = 0,
\cr
\sp(3.0)} \eqno(20)$$
$$ \eqalign{ \sp(2.0)
\nabla^2 f = 0.
\cr
\sp(3.0)} \eqno(21)$$
Since we are interested in the physics only in the vicinity of the 
apparent horizon, it is sufficient to verify that the Vaidya metric (16) 
and the dilaton field (17) are consistent with the field equations 
(19)-(21) near the apparent horizon. After some manipulation, the field 
equations require to satisfy
$$ \eqalign{ \sp(2.0)
\partial_r f = \partial_v \partial_r f \approx 0,
\cr
\sp(3.0)} \eqno(22)$$
$$ \eqalign{ \sp(2.0)
\partial_v M \approx {1 \over 2} (\partial_v f)^2 ,
\cr
\sp(3.0)} \eqno(23)$$
where we shall use $\approx$ to express the equalities holding
approximately near the apparent horizons from now on. From (22) and (23), 
we find the general solution
$$ \eqalign{ \sp(2.0)
f(v) \approx  \pm \int^v dv \sqrt{2 \partial_v M}.
\cr
\sp(3.0)} \eqno(24)$$
Consequently, it has been checked that the solutions (16) and (17) are at least 
classically consistent with the field equations near the apparent horizon 
as long as (24) is satisfied. Incidentally, (24) represents the physical 
fact that the increase of the black hole mass, $\partial_v M > 0$, is 
classically allowed, but the loss of it, $\partial_v M < 0$, i.e., the 
Hawking radiation, is classically forbidden and can occur only through 
the quantum tunneling effects owing to $f(v)$ being the real scalar field.

\vskip 32pt
\leftline{\bf 4. Quantum Gravity near Apparent Horizon}	
\par
We now consider the dynamical black hole (16) and the linear dilaton 
(17). Our main concern in this section is to construct a quantum theory 
of two dimensional dilaton gravity holding near the apparent horizon. 

Let us begin by introducing the coordinates
$$ \eqalign{ \sp(2.0)
x^a = (x^0, x^1) = (v - r, r).
\cr
\sp(3.0)} \eqno(25)$$
Next we set up the gauge conditions such that the gauge symmetries 
associated with the two dimensional reparametrization invariances are 
completely fixed
$$ \eqalign{ \sp(2.0)
g_{ab} &= \left(\matrix{ { - \alpha^2 + {\beta^2 \over \gamma}} & \beta \cr
              \beta & \gamma \cr} \right),
\cr
 &= \left(\matrix{ -(1 - {M \over \lambda} e^{-2 \lambda r})  & {M \over 
                   \lambda} e^{-2 \lambda r}  \cr
                   {M \over \lambda} e^{-2 \lambda r} & 1 + {M \over 
                   \lambda} e^{-2 \lambda r}} \right) ,
\cr
\sp(3.0)} \eqno(26)$$
where the black hole mass $M$ is the function of the two dimensional 
coordinates $x^a$.  Notice that we have chosen these gauge conditions to 
correspond to the Vaidya metric built in the previous section. Of course, 
at this stage, we cannot restrict the mass function to be the function 
depending on only the $v$ coordinate from an argument of symmetries. 
Near the apparent horizons (18), (26) yields
$$ \eqalign{ \sp(2.0)
\alpha \approx {1 \over \sqrt{2}} , \ \beta \approx 1, \ \gamma = {1 
\over \alpha^2} \approx 2.
\cr
\sp(3.0)} \eqno(27)$$
Note that the dynamical degrees of freedom representing ``graviton'' 
$\gamma$  are effectively fixed in (27). At this point, let us make 
physically plausible assumptions [4, 5] near the apparent horizon  
$$ \eqalign{ \sp(2.0)
f \approx f(v), \ M \approx M(v) , \ \phi \approx - \lambda r. 
\cr
\sp(3.0)} \eqno(28)$$
As shown in the section 3, these assumptions are consistent with the 
field equations, but their quantum-mechanical meaning is not clear at 
present. Given the assumptions (28), near the apparent horizon the 
canonical conjugate momenta (8) become
$$ \eqalign{ \sp(2.0)
p_f &\approx  \partial_v f,
\cr
p_{\phi} &\approx  - 2 M - {1 \over \lambda} \partial_v M,
\cr
p_{\gamma} &\approx  M.
\cr
\sp(3.0)} \eqno(29)$$
Then after a little lengthy calculation, one arrives at a remarkable 
relation that the Hamiltonian constraint $H_0 = 0$ becomes 
proportional to the supermomentum constraint $H_1 = 0$
$$ \eqalign{ \sp(2.0)
\sqrt{2} H_0 &\approx 2  H_1,
\cr
             &\approx p_f ^2  + 2 \lambda p_\phi + 4 \lambda M.
\cr
\sp(3.0)} \eqno(30)$$
This relation can be understood from an observation that the time 
translation generated by the Hamiltonian constraint is frozen on the 
apparent horizon due to gravitational time dilation in the present   
coordinate system [6]. 

We are now ready to carry out the canonical quantization of the model. 
Following Dirac's quantization procedure of the first-class constraints 
[12], the residual symmetry (30) is imposed on the state
$$ \eqalign{ \sp(2.0)
( - {\partial^2 \over {\partial f^2}} - 2 i \lambda {\partial 
\over {\partial \phi}} + 4 \lambda M ) \Psi = 0,  
\cr
\sp(3.0)} \eqno(31)$$
which is nothing but the Wheeler-DeWitt equation. A special solution can 
be found to be 
$$ \eqalign{ \sp(2.0)
\Psi = (  B e^{\sqrt{A} f(v)} + C e^{-\sqrt{A} f(v)} ) e^{ i {{ A - 4 
\lambda M } \over {2 \lambda}} \phi },
\cr
\sp(3.0)} \eqno(32)$$
where $A$, $B$, and $C$ are integration constants. Without losing  
generality, we shall choose the boundary condition $B = 0$.

\vskip 32pt
\leftline{\bf 5. Hawking Radiation}	
\par
We now turn our attention to an application of the quantum gravity near the 
apparent horizon for understanding the Hawking radiation in the two 
dimensional dilaton gravity.  A similar analysis was carried out in four 
[4, 5] and three [6] dimensional black holes.  The main motivation behind the 
present work is to clarify to what extent the quantum gravity holding near 
the apparent horizon reflects the physical properties of quantum black 
holes since we have a better grasp of the quantum mechanical features of 
a black hole in the two dimensional dilaton gravity compared to in the 
four and three dimensional gravities.

To begin with, let us define the expectation value $< \cal O >$ of an 
operator  $\cal O$ by
$$ \eqalign{ \sp(2.0)
< {\cal O} > = {1 \over {\int df \ |\Psi|^2}}  \int df \ \Psi^{\dag} {\cal 
O} \Psi.
\cr
\sp(3.0)} \eqno(33)$$
Under this definition, it is straightforward to evaluate the expectation 
value of the change rate in black hole mass
$$ \eqalign{ \sp(2.0)
< \partial_v M > = - {A \over 2},
\cr
\sp(3.0)} \eqno(34)$$
where the constraint (30) (or $p_{\phi}$ in (29)) and the physical state 
(32) were used. Moreover, in a similar manner one can calculate that in 
the radius in the apparent horizon
$$ \eqalign{ \sp(2.0)
< \partial_v r_{AH} > = - {A \over {4 \lambda < M >}}.
\cr
\sp(3.0)} \eqno(35)$$

To represent the Hawking radiation, we have to select the integration 
constant $A$ to be a positive constant, for example, $k_1 ^2$, then (32), 
(34) and (35) reduce to 
$$ \eqalign{ \sp(2.0)
\Psi = C \ e^{- | k_1 | f(v) + i {{ A - 4 
\lambda M } \over {2 \lambda}} \phi }.
\cr
\sp(3.0)} \eqno(36)$$
$$ \eqalign{ \sp(2.0)
< \partial_v M > = - {k_1 ^2 \over 2},
\cr
\sp(3.0)} \eqno(37)$$
$$ \eqalign{ \sp(2.0)
< \partial_v r_{AH} > = - {k_1 ^2 \over {4 \lambda < M >}}.
\cr
\sp(3.0)} \eqno(38)$$
To see explicitly that this is in fact the Hawking radiation carried by 
the matter field $f$, it is useful to argue the expectation value of the 
energy-momentum tensor of the matter field, which is defined as
$$ \eqalign{ \sp(2.0)
< T^f \ _{ab} > &= < {1  \over \sqrt{-g}}{\delta S_f \over \delta 
g^{ab}} >,  
\cr
&= - {1 \over 2} < \nabla_a f  \nabla_b f - {1 \over 2} g_{ab} ( 
\nabla f)^2 >,
\cr
\sp(3.0)} \eqno(39)$$
where $S_f$ is the matter part in the action (1). Then $T_{vv}$ in which 
we are interested is calculated to be
$$ \eqalign{ \sp(2.0)
< T^f \ _{vv} > &= {k_1 ^2 \over 2},
\cr
\sp(3.0)} \eqno(40)$$
which is precisely equal to the opposite sign of (37), thus means that 
the matter flux is equivalent to the Hawking radiation as expected. 
Alternatively, if we choose $A$ to be a negative constant, 
e.g., $- k_2 ^2$, we gain a physical situation where external neutral 
matters flow into a black hole across the horizon. 

Several comments about (36)-(38) are now in order. Firstly, (36) shows 
that the physical state has an exponentially damping-like form in the 
classically forbidden region implying the quantum tunneling process. This 
behavior seems to match our interpretation of the present situation as 
the Hawking radiation. Secondly, from (37), the Hawking radiation rate is 
independent of the black hole mass in contrast with the case of the four 
dimensional Schwarzschild black hole where it is shown that $< \partial_v 
M > \propto - {1 \over  < M^2 >}$ [4, 5] inferred by Hawking in his 
semiclassical approach [1]. This surprising result, however, has been 
already found in studies of the two dimensional dilaton and the other two 
dimensional gravities [13]. This is because in two dimensions there exists a 
beautiful relation between the trace anomaly and the Hawking radiation 
[14]. Namely, although for $N$ species of massless scalar field the 
trace of the energy-momentum tensor vanishes classically, 
quantum-mechanically there is the trace anomaly 
$$ \eqalign{ \sp(2.0)
< g^{ab} T^f \  _{ab} > =  {N \over 24} \ R.
\cr
\sp(3.0)} \eqno(41)$$
(Up to now we take account of the case $N = 1$.) As a 
result, in the coordinates where the metric is asymptotically constant on 
the null infinities $\cal I_R ^{\pm}$, the Hawking radiation rate is 
asymptotically independent of the black hole mass and approaches the 
constant value ${N \lambda^2 \over 48}$ [8]. One  might ask what would 
happen if one replaces a single scalar matter $f$ with $N$ species of 
scalar $f_i \ (i=1, 2,...,N)$ in our formalism as considered in the CGHS  
original work. In this case, the physical state (32) is replaced by
$$ \eqalign{ \sp(2.0)
\Psi = \prod_{i=1} ^{N} (  B_i e^{\sqrt{A_i} f_i(v)} + C_i e^{-\sqrt{A_i} 
f_i(v)} ) \ e^{ i {{ A - 4 
\lambda M } \over {2 \lambda}} \phi },
\cr
\sp(3.0)} \eqno(42)$$
with
$$ \eqalign{ \sp(2.0)
\sum_{i=1} ^{N} A_i = A,
\cr
\sp(3.0)} \eqno(43)$$
and the result with respect to the Hawking 
radiation (34) remains unchanged. If we consider completely identical $N$ 
scalar matters such that $A = N \bar A, (A_1 = A_2 = .....= A_N \equiv 
\bar A)$, the Hawking radiation rate becomes to scale with $N$
$$ \eqalign{ \sp(2.0)
< \partial_v M > = - {N \bar A \over 2},
\cr
\sp(3.0)} \eqno(44)$$
which is equal to the result derived by CGHS up to a numerical constant. 

Finally, (38) shows that as a black hole emits the Hawking radiation and 
loses the mass the apparent horizon recedes toward the singularity. This 
is a very plausible picture from physical consideration. In the limit 
$< M > \rightarrow 0$, this equation indicates that $< \partial_v r_{AH} 
> \rightarrow - \infty$, suggesting that the apparent horizon disappears 
and the curvature singularity might be visible to external observers 
(violation of weak cosmic censorship). However, in order to get a 
definite answer to this problem, it seems that we need a more improved 
and sophisticated model as discussed in the next section.

Before closing this section, it is valuable to inquire what would happen 
near the  
origin $r = 0$. To keep the role of $x^0$ coordinate as the time, we here 
assume the inequality $M < \lambda$. Surprisingly enough, despite $r = 
0$ being the curvature singularity [15], it turns out that one can also 
establish quantum gravity in this vicinity in a perfectly similar way to 
the case of the apparent horizon. For example, 
analogous equations to (27), (29), (30), (32) and (34) can be deduced as 
follows: 
$$ \eqalign{ \sp(2.0)
\alpha \approx {1 \over \sqrt{1 + {M \over \lambda}}} , \ \beta \approx 
{M \over \lambda}, \ \gamma = {1 \over \alpha^2} \approx 1 + {M \over 
\lambda}, 
\cr
\sp(3.0)} \eqno(45)$$
$$ \eqalign{ \sp(2.0)
p_f \approx  \partial_v f, \ p_{\phi} \approx  - {2 \over \gamma 
\lambda} ( \partial_v M + 2 M^2 ), \ p_{\gamma} \approx  {2 M \over 
\gamma}, 
\cr
\sp(3.0)} \eqno(46)$$
$$ \eqalign{ \sp(2.0)
\sqrt \gamma H_0 &\approx \gamma  H_1,
\cr
             &\approx p_f ^2  +  \lambda \gamma p_\phi + 4 M^2,
\cr
\sp(3.0)} \eqno(47)$$
$$ \eqalign{ \sp(2.0)
\Psi = (  B e^{\sqrt{A} f(v)} + C e^{-\sqrt{A} f(v)} ) e^{ i {A - 4 M^2 
\over {\lambda \gamma}} \phi},
\cr
\sp(3.0)} \eqno(48)$$
$$ \eqalign{ \sp(2.0)
< \partial_v M > = - {A \over 2}.
\cr
\sp(3.0)} \eqno(49)$$
These results, in particular, (49), are obviously consistent with results 
obtained by means of  
quantum gravity holding near the apparent horizon. One interesting 
problem in future would be to find the physical state holding in the 
region between the curvature singularity and the apparent horizon by 
connecting the two states (32) and (49) consistently.

\vskip 32pt
\leftline{\bf 6. Conclusion}	
\par
So far we have investigated the Hawking radiation in terms of quantum 
gravity holding near the apparent horizon. In particular, it was found 
that the Hawking radiation rate is independent of the black hole mass, 
and it scales with the number of massless scalar fields when there are 
$N$ identical matters as shown in the other analysis of the two 
dimensional dilaton gravity. 

Since the main motivation in the present study is to compare the results 
obtained in the quantum gravity holding near the apparent horizon with 
those of fully quantized two dimensional dilaton gravity, we should 
make comments on their relationship more closely. First of all, we notice 
that the value of the coefficient of the Hawking flux cannot be determined 
in the present formalism treating only the region in the vicinity of the 
horizon since it is fixed by imposing the boundary conditions at the null 
past infinities $\cal I_R ^-$ and $\cal I_L ^-$ [8, 13]. Nevertheless, it 
is  
remarkable that the present formalism describes qualitative features of 
the Hawking radiation without paying much attention to the interior and 
the exterior regions of a black hole. Actually, the quantum 
gravity near the horizon has provided a nice description of the Hawking 
radiation in three dimensional de Sitter black hole [16] where there is no 
asymptotically flat regions so that the usual technique cannot be applied to 
this case [6].

Next as mentioned before, it was known that in two dimensions the Hawking 
radiation stems from the trace anomaly [14]. Since the present formalism 
gives  
the qualitatively same picture as the semiclassical approach, it is 
certain that our formalism respects the contribution from the trace 
anomaly properly. However, it is conjectured that the functional measures 
over the gravitational field, the dilaton and the ghosts give rise to the 
nontrivial contributions which are different from the form of the trace 
anomaly, and eventually yield the backreaction of the Hawking radiation 
on the  geometry although we have not yet reached the fully consistent 
model even in the two dimensional dilaton gravity [13]. In this respect, 
unfortunately it seems that our formalism largely ignores this issue. But 
this problem is shared in the more general ADM  or  the Wheeler-DeWitt 
formalisms where we have no idea how to evaluate the functional measures. 
It is likely that one should set up the Wheeler-DeWitt equation after 
estimating the contributions from the functional measures carefully 
though we do not know how to accomplish this procedure except in two 
dimensions. Although we have a lot of things to overcome in the future, 
we believe that an improvement of the formalism at hand would give us a 
useful analytical method for understanding various properties of quantum 
black holes and might to a certain extent realize the Membrane Paradigm 
of a black hole [17-19].

\vskip 32pt
\leftline{\bf References}
\centerline{ } %
\par
\item{[1]} S.W.Hawking, Comm. Math. Phys. {\bf 43} (1975) 199.

\item{[2]} D.Page, in Proceedings of the Fifth Canadian Conference on 
General Relativity and Relativistic Astrophysics, edited by 
R.B.Mann et al. (World Scientific, Singapore, 1994). 

\item{[3]} J.D.Bekenstein, Phys. Rev. {\bf D7} (1973) 2333; ``Do We 
Understand Black Hole Entropy?'', gr-qc/9409015; R.M.Wald, ``Black Holes 
and Thermodynamics'', gr-qc/9702022. 

\item{[4]} A.Tomimatsu, Phys. Lett. {\bf B289} (1992) 283.

\item{[5]} A.Hosoya and I.Oda, Prog. Theor. Phys. {\bf 97} (1997) 233.

\item{[6]} I.Oda, ``Evaporation of Three Dimensional Black Hole in 
Quantum Gravity'', EDO-EP-9, gr-qc/9703055; ``Quantum Instability of 
Black Hole Singularity in Three Dimensions'', EDO-EP-10, gr-qc/9703056.

\item{[7]} I.Oda, ``Mass Inflation in Quantum Gravity'', EDO-EP-8, 
gr-qc/9701058; ``Cosmic Censorship in Quantum Gravity'', EDO-EP-11, 
gr-qc/9704021.

\item{[8]} C.G.Callan, S.B.Giddings, J.A.Harvey and A.Strominger, Phys. 
Rev. {\bf D45} (1992) R1005.

\item{[9]} P.Hajicek, Phys. Rev. {\bf D30} (1984) 1178; P.Thomi, B.Isaak  
and P.Hajicek, Phys.  Rev. {\bf D30} (1984) 1168.

\item{[10]} R.M.Wald, General Relativity (The University of Chicago Press, 
1984).

\item{[11]} T.Regge and C.Teitelboim, Ann. Phys. (N.Y.) {\bf 88} (1974) 286.

\item{[12]} P.A.M.Dirac, Lectures on Quantum Mechanics (Yeshiva University, 
1964). 

\item{[13]} J.A.Harvey and A.Strominger, ``Quantum Aspects of Black 
Holes'', hep-th/9209055.

\item{[14]} S.M.Christensen and S.A.Fulling, Phys. Rev. {\bf D15} (1977) 
2088.

\item{[15]} E.Witten, Phys. Rev. {\bf D44} (1991) 314.

\item{[16]} M.Ba$\tilde n$ados, C.Teitelboim, and J.Zanelli, Phys. Rev. 
Lett. {\bf 69} (1992) 1849; M.Ba$\tilde n$ados, M.Henneaux, C.Teitelboim, 
and J.Zanelli, Phys. Rev. {\bf D48} (1993) 1506.

\item{[17]} K.S.Thorne, R.H.Price and D.A.Macdonald, Black Hole: The 
Membrane Paradigm (Yale University Press, 1986).

\item{[18]} G.'t Hooft, Nucl. Phys. {\bf B335} (1990) 138; Phys. Scripta 
{\bf T15} (1987) 143; ibid. {\bf T36} (1991) 247.

\item{[19]} I.Oda, Int. J. Mod. Phys. {\bf D1} (1992) 355; Phys. Lett. 
{\bf B338} (1994) 165; Mod. Phys. Lett. {\bf A10} (1995) 2775.

\endpage
%

%
\bye